\documentclass[aps,pre,twocolumn,floatfix]{revtex4} 
\usepackage{amsmath}
\usepackage{amssymb}   
\usepackage{graphicx}
\usepackage{revsymb}

\begin{document}

\title{Vortices in vibrated granular rods}

\author{Daniel L. Blair,  T. Neicu and A. Kudrolli}

\affiliation{Department of Physics, Clark University, Worcester, MA
01610, USA}

\date{\today}

\begin{abstract}
We report the experimental observation of novel vortex patterns in
vertically vibrated granular rods. Above a critical packing fraction,
moving ordered domains of nearly vertical rods spontaneously form and
coexist with horizontal rods.  The domains of vertical rods coarsen in
time to form large vortices.  We investigate the conditions under
which the vortices occur by varying the number of rods, vibration
amplitude and frequency. The size of the vortices increases with the
number of rods. We characterize the growth of the ordered domains by
measuring the area fraction of the ordered regions as a function of
time.   A {\em void filling} model is presented to describe the
nucleation and growth of the vertical domains.  We track the ends of
the vertical rods and obtain the velocity fields of the vortices. The
rotation speed of the rods is observed to depend on the vibration
velocity of the container and on the packing.  To investigate the
impact of the direction of driving on the observed phenomena, we
performed experiments with the container  vibrated horizontally.
Although vertical domains form, vortices are not observed.   We
therefore argue that the motion is generated due to the interaction of
the inclination of the rods with the bottom of a vertically vibrated
container.  We also perform simple experiments with a single row of
rods in an annulus.  These experiments directly demonstrate that the
rod motion is generated when the rods are inclined from the vertical,
and is always in the direction of the inclination.

\end{abstract}

\maketitle

\section{Introduction}
Granular materials are well known examples of dissipative
nonequilibrium systems that show a rich variety of collective
phenomenon such as convection, wave patterns, and
segregation~\cite{Melo94,knight93,tennakoon98,makse97}.   Most studies
utilize spherical particles to investigate these bulk properties.
However, in most natural or industrial settings one can find 
an abundance of anisotropic granular materials {\em viz.}~rice,
medicine capsules, and even logs.  Therefore it is surprising that
very few studies of prolate granular materials have been carried out. 
Mounfield and Edwards~\cite{Mounfield} applied the concepts of configurational
statistical mechanics to study the nature of the isotropic to nematic
phase transition in a granular system of elongated particles. In
recent experiments utilizing a tall narrow cylinder, Villarruel {\it
et al.}~\cite{Villarruel} studied the effects of anisotropy on
granular packing.  They observed the appearance of smectic states with
the direction given by the container walls. 

In thermal systems, particle anisotropy is known to produce ordered
states. Examples are rod and plate shaped colloids and liquid
crystals, which show orientational order and form nematic and smectic
phases~\cite{Onsager,Khokhlov,Dogic,Kooij,deGennes}. The ordering
mechanism was found to be entropically driven, {\em i.e.}~as
thermodynamic equilibrium is reached, the process of entropy
maximization leads to long range order. It is not obvious that such
mechanisms carry over to granular systems because the thermal energy
scale is very much smaller than the potential energy required for
particle rearrangement, and energy has to be supplied actively to
produce sustained motion.  Therefore, an interesting question arises --
does shape anisotropy lead to self-organization and pattern formation
in granular materials?

In this paper, we report the observation of novel vortex patterns
exhibited by granular rods that are vibrated vertically inside a
container.  We obtain the phase diagram for the observed patterns as a
function of the acceleration of the driving and the packing fraction
of the rods.  We find that for sufficiently large packing fractions,
the rods tend to align vertically and undergo vortex motion. Using
high frame rate imaging and particle tracking, we have measured the
velocity fields of the vortices as a function of packing fraction and
driving frequency.  Based on our experimental  observations, we argue
that the inclination of the rods causes motion due to collisions with
the bottom boundary. To bolster our claim, we conduct experiments with
single row of rods in an annulus to demonstrate that motion is always in the
direction of inclination. Thus we show that shape anisotropy leads to
translational motion in such systems.

\section{Experimental apparatus}
The experiments were performed in a circular anodized aluminum
container with a diameter $D = 6.0$ cm and depth $H = 1.5$ cm. (Limited
experiments were also carried out in a container with $D = 8.5$ cm.)
The container is leveled to within 0.002 cm and is attached to an
electro-mechanical shaker through a rigid linear bearing that allows
motion only in the vertical direction. The cell is driven by a
sinusoidal signal at a frequency $f$, and is monitored by an
accelerometer via a lock-in amplifier. The experiments were conducted
using copper cylinders with uniform length $l= 6.2$ mm and diameter $d
= 0.5$ mm. The cylindrical surface of the rods is coated with a grey
tin-oxide layer to diminish light reflection and has a coefficient of
friction $\mu = 0.32$. The patterns that form are imaged from above
using a high-frame rate digital camera (Kodak SR-1000). Since the flat
ends of the rods reflect light better than the  sides, they appear as
bright spots when imaged from above. If the rods are inclined greater
than 35 degrees, they reflect far less light compared to nearly
vertical and horizontal rods.

One control parameter for our experiments is $\Gamma = {\cal A}(2 \pi
f)^2 /g$, where ${\cal A}$ is the driving amplitude and $g$ is the
acceleration  due to gravity.  A second control parameter is the
non-dimensional number fraction  $ n_\phi
= N / N_{max}$, where $N$ is the total number of rods in the container
and $N_{max}= \frac{\pi}{\sqrt{12}} \left( \frac{D}{d} \right )^2$ is
the maximum number of rods required to obtain a vertically aligned
monolayer. Therefore, $n_\phi =1$ corresponds to having one layer of
triangularly packed vertical rods.

\section{\label{obs}Observed patterns}
\begin{figure}[t]
\includegraphics[width=0.45\textwidth]{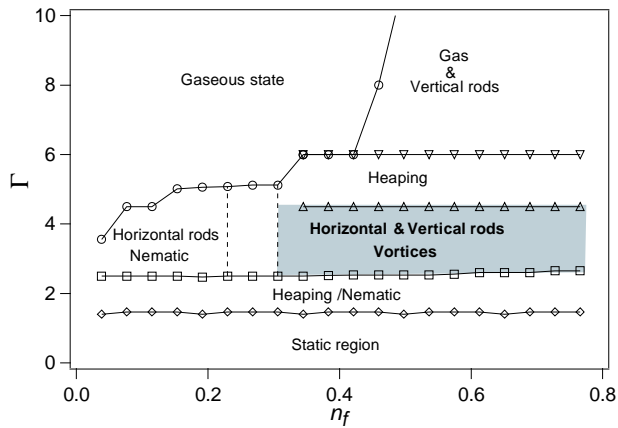}
\caption{\label{phase-dia}The phase diagram of the observed patterns
in vibrated granular rods. The driving frequency $f = 50$ Hz. Vortices are
observed for sufficiently large numbers of rods and driving
acceleration parameter $\Gamma$.}
\end{figure}

\begin{figure}[t]
\includegraphics[width=0.45\textwidth]{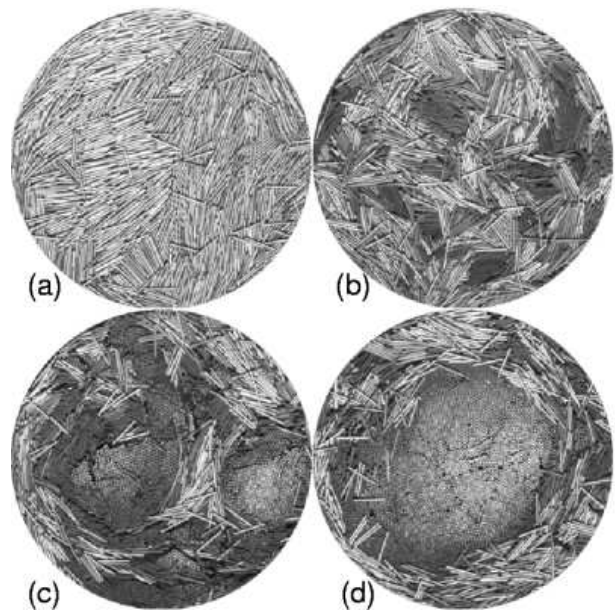}
\caption{\label{vort_time} Examples of patterns observed when granular rods
are vertically vibrated $(f = 50\, \rm{Hz})$. (a) Nematic-like state
($n_{\phi} = 0.152$, $\Gamma = 2.42$). (b) Moving domains of near
vertical rods ($n_{\phi} = 0.344$, $\Gamma = 3.38$). (c) Multiple
vortices ($n_{\phi} = 0.551$, $\Gamma = 3.29$).  (d) Large vortex
($n_{\phi} = 0.535$, $\Gamma = 3.00$). The cylindrical surface of the
rods is coated with a grey tin-oxide layer to diminish light
reflection and has a coefficient of friction $\mu = 0.32$. The
patterns that form are imaged from above using a high-frame rate
digital camera (Kodak SR-1000). Since the tips 	of the rods
reflect light better than the  sides, they appear as bright spots when
imaged from above. If the rods  are inclined greater than 35 degrees,
they reflect far less light compared  to nearly vertical and
horizontal rods.}
\end{figure}

The system is initialized by pouring the rods into the container and
then increasing $\Gamma$, which establishes a random
state. Figure~\ref{phase-dia} shows the phase diagram of the observed
patterns as a function of $n_\phi$ ($f = 50$ Hz). The rods are
observed to remain static in the initial configuration until $\Gamma
\ge 1.47$. As $\Gamma$ is increased above this value, the rods heap
towards one side of the container similar to previous observations
with spherical particles~\cite{heap-ref}. As the acceleration is
increased further above $\Gamma = 2.6$, the rods are observed to
spread-out evenly inside the container, and form horizontal layers
with random orientations. An example of such a nematic phase observed
at low $n_\phi$ is shown in Fig.~\ref{vort_time}(a). Above a
critical $n_\phi$, a second transition is observed. Chaotically moving
domains of almost vertical rods are observed to spontaneously form and
coexist with horizontal rods~\cite{fraden}. An example is shown in
Fig.~\ref{vort_time}(b).

As $n_\phi$ is increased further, the domains of near-vertical
rods that are formed coalesce and undergo {\em vortex motion}.
Fig.~\ref{vort_time}(c) shows an example of two counter-rotating
vortices. For large number fractions ($n_\phi > 0.49$), a single
large stable vortex made entirely of near-vertical rods in the center
and inclined rods at its' boundary, is observed
[Fig.~\ref{vort_time}(d)].

If the acceleration is increased further, a gas like state is reached
where the rods vibrate vigorously inside the container and the nematic
and vortex states are destroyed. For the highest $n_\phi$,  a pure
gaseous state is not reached due to the limits of the apparatus. 
Instead, domains of ordered vertical rods are observed to coexist with
the gas of rods. 

\subsection{\label{growth}Domain Growth}

\begin{figure}[t]
\includegraphics[width=0.45\textwidth]{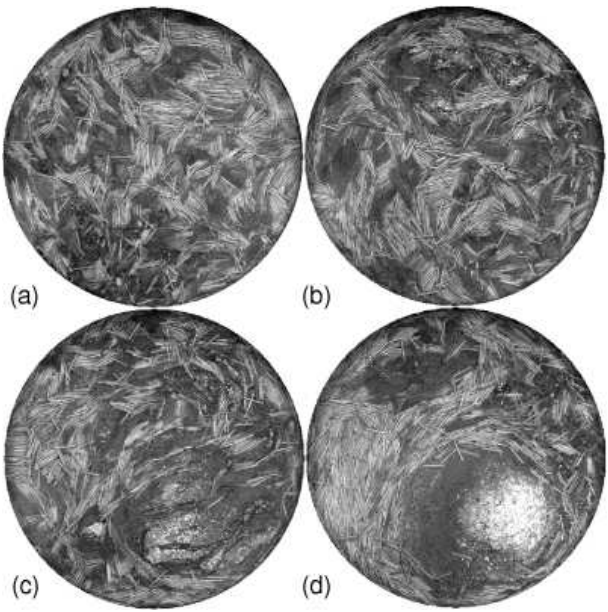}
\caption{\label{vort_t}The formation of the vortex from an initial
random state as function of time $t$ of vibration with $\Gamma = 2.6$
and $n_{\phi}= 0.39$. (a) t = 3 min, (b) t
= 6 min, (c) t = 9 min, (d) t = 12 min. Small domains of vertical rods
are observed to form and coarsen over time and evolve into a
vortex. ($f =$ 60 Hz).}  
\end{figure}

We now discuss how vortices nucleate and grow as a function of time.
An example of the growth process is shown in Fig.~\ref{vort_t}, and
a movie can be viewed at Ref.~\cite{movies}.  Starting from a
random state, pockets of near-vertical domains are observed to
nucleate uniformly inside the container. This is in contrast with previous
observations where the ordered domains grow inward from the
boundary~\cite{Villarruel}. The pockets of vertical domains grow in time
by merging to form larger domains of almost vertical rods.  The
domains are then observed to collectively move as the the system becomes more
ordered and then finally show vorticity.

\begin{figure}[t]
\includegraphics[width=0.45\textwidth]{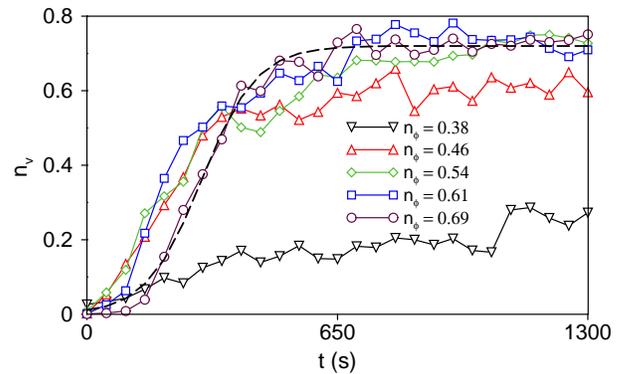}
\caption{\label{coarsen}The fraction of vertical rods $n_v$ as a
function of time for various $n_{\phi}$ at $\Gamma=3.4$ and $f=50$ Hz.
The vertical domains show two distinct growth regimes.  
At short times, the growth is increasing exponentially and then
gives way to an exponential decrease in growth to a saturation value
determined by $n_{\phi}$.  For $n_{\phi} = 0.38$, the domain growth
saturates very slowly to a value less than $n_\phi$ because the near
vertical domains cannot be supported by the horizontal rods.  
For $n_{\phi} \ge 0.46$ the two distinct growth regimes are apparent
and for all $n_{\phi}$ vortex motion occurs after $t=300$ s. We note
that for the highest value of $n_{\phi}$ the initial growth displays a
slow down, which is directly related to the void filling mechanism
discussed in the text.  The dashed line is a fit to the integrated
growth equation discussed in Sec.~\ref{discuss}.  We find that the
simplified model gives an accurate interpretation of the results.}
\end{figure}

We measure the growth of the domains by taking time-lapse images of
the process described above.  The growth of near-vertical domains is
then measured from the ratio of the domain area to that of the total
container area as a function of time.   The fraction of rods that are
near-vertical, $n_v$ is plotted in Fig.~\ref{coarsen} for a range of
$n_{\phi}$ at a fixed driving acceleration and frequency,  $\Gamma = 3.4$ and
$f=50$ Hz.  

We find that there exist two growth regimes [a model is discussed in
Sec.~\ref{discuss}]. For short times, the growth rate depends on
$n_{\phi}$ in the following way.  If $n_{\phi}$ is below a critical
value, domains appear but never organize into large vortices.
However, as $n_{\phi}$ is increased the short time growth is slowed by
the high packing.  That is, more time is required for small domains to
form due to the lack of {\em voids} present.  A void filling model
for domain growth will be discussed in Sec.~\ref{discuss}. The
saturation value of $n_v$ increases with $n_{\phi}$.  The difference in
the values arises due to the definitions of $n_v$ and $n_{\phi}$.

\begin{figure}[t]
\includegraphics[width=0.45\textwidth]{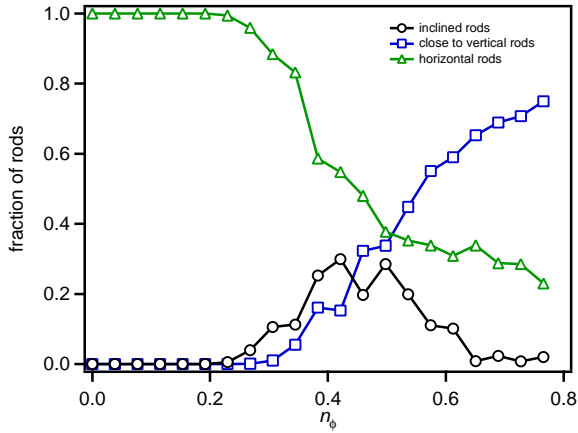}
\caption{\label{v-h-frac} The relative number of inclined, vertical
and horizontal rods as a function of $n_\phi$ after steady state is
reached $\Gamma = 3.00$ and $f = 50$ Hz. }
\end{figure}

The variation of the relative number of near-vertical rods as a
function of $n_\phi$ is plotted in Fig.~\ref{v-h-frac}. We show
separate data for the relative number of inclined rods, vertical rods
and horizontal rods.  Rods that are tilted between 30 and 80 degrees
are considered inclined, and the rest are considered as either
horizontal or vertical (the inclination of the rod was obtained by
measuring the distance $x$ between adjacent rods and calculating
sin$^{-1}[d/x]$). For $n_\phi < 0.22$, the container has only
horizontal rods. As $n_\phi$ increases, more rods tend to the vertical
direction. For $n_\phi >$ 0.71 most rods are either vertical or
horizontal.

\subsection{Spatial structure of the vortex and velocity fields}

\begin{figure}[t]
\includegraphics[width=0.425\textwidth]{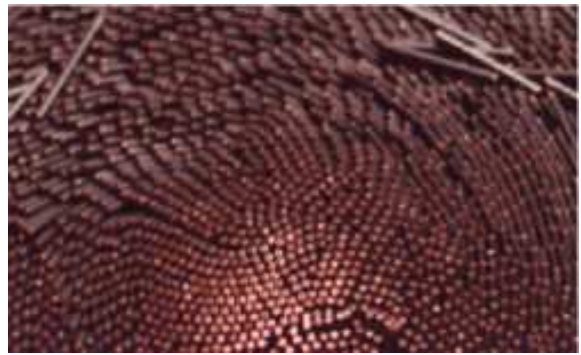}
\caption{ \label{vortex} Close-up image of a vortex that has self
organized at $n_{\phi} = 0.419$ (only the upper half of the vortex
is shown).  The rods are inclined in the direction of rotation, which
is counterclockwise.  The inclination angle of the rods varies from
about 90 degrees at the center of the vortex to near 40 degrees close
to the  edge.}
\end{figure}

We next discuss the spatial structure of the vortex. In
Fig.~\ref{vortex}, a close-up image of a vortex pattern is displayed
which shows the progressive inclination of the rods from the center of
the vortex.  We note that the direction of motion corresponds to the
direction of inclination (for Fig.~\ref{vortex} the motion is
counterclockwise).  In  this case
the inclination of the rods varies from approximately 85 to 40
degrees near the edges. The change in inclination of the rods within
the vortex is less at higher $n_\phi$ due to the increased packing.

\begin{figure}[t]
\includegraphics[width=0.45\textwidth]{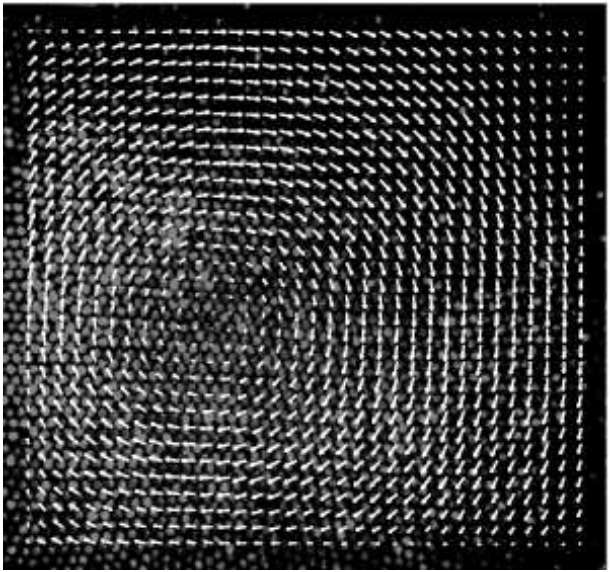}
\caption{\label{field}The velocity field of a granular vortex
($n_{\phi} =  0.497, \Gamma = 3.00$). The data was obtained by
tracking individual rod ends and then by spatial and time averaging
the obtained displacements (see text).}
\end{figure}

By acquiring images at 250 (frames s$^{-1}$), we were able to track
the ends of the vertical rods and obtain the velocity fields of the
vortices.  In addition to the rotational motion, rods also vibrated
due to the collisonal interactions with their neighbors.  We first
track~\cite{IDL} the ends of the near-vertical rods and then perform
spatial and temporal averaging to obtain the vortex fields. An
example of the velocity field of a vortex is presented in
Fig.~\ref{field} (here the temporal averages include 100 consecutive
frames, and spatial averages were taken over an area of $2d \times 2d$).

The averaged azimuthal velocity $v(r)$, as a function of the distance
$r$ from the center of the vortex, is shown in Fig.~\ref{Fig4}(a) for
a range of $n_\phi$. The data was obtained at a constant acceleration
$\Gamma = 3.00$ for a frequency $f = 50$ Hz. We note that for small
distances relative to the center of the vortices, the averaged
velocity $v(r)$ increases linearly with the distance $r$ indicating
solid body rotation. At intermediate ranges, this linear relationship
is not observed, and therefore shearing occurs inside the vortices. As
the boundary of the vortex is approached, velocity decreases due to
friction with the horizontal rods at the edge. It is interesting  to
note that the horizontal rods at the edge of the vortex are often
aligned tangentially with the boundary of the vortex [see
Fig.~\ref{vort_time}(c)]. We also observe that the slope of the
averaged velocity $v(r)$ at small $r$ systematically decreases as the
packing fraction $n_\phi$ is increased. Thus the angular velocity of
the inner core of the vortex decreases with its size.

\subsection{Frequency dependence}

We also explored the rotation rate dependence of the vortices as a
function of $f$.  Figure~\ref{Fig4}(b) shows $v(r)$ for $n_\phi = 0.53$ and
$\Gamma = 3.00$ for vortices of equivalent size.  As stated, $\Gamma$
is held constant while frequency is increased, which decreases
vibration velocity.  Therefore, we demonstrate that the vortex speed
systematically increases with vibration velocity.

\section{Horizontally vibrated container}
To test if the transition to a vertical state is dependent on the way
the container is vibrated, experiments were performed with {\em
horizontal} driving. There do in fact exist similarities: at low
$n_\phi$ heaping and nematic like domains of horizontal rods are
observed; at high $n_{\phi}$ the rods tend to align vertically and
form ordered domains. An image is shown in Fig.~\ref{horizontal}. In
contrast to the vertical shaking the vertically aligned domains do not
migrate and coarsen into vortices. We also find that in this case the
convection is stronger and destroys the motion of the vertically
ordered domains.

\section{\label{discuss}Discussion}

We first discuss possible mechanisms responsible for the rod to align
vertically at high packing fractions. The tendency of the rods to
align vertically at high packing fractions may be understood in terms
of a void filling mechanism.  If voids ({\em i.e.}~space between rods), are
small then they can only be filled with a near-vertical rod.  On the
other hand, larger voids will accommodate a horizontal rod.  Assuming
that the distribution of voids decreases with void size, the most
probable configuration will be regions of vertical rods.  Furthermore,
the decrease in the number of large voids is enhanced by the
coalescence of regions of vertically aligned rods.  This also drives
the system to pack more closely and produces a decreased overall
center of mass. Thus ordering at high packing
fractions can be explained by a process that is analogous to configurational
entropy driven ordering in thermal systems~\cite{Mounfield}.

\begin{figure}[t] 
\includegraphics[width=0.45\textwidth]{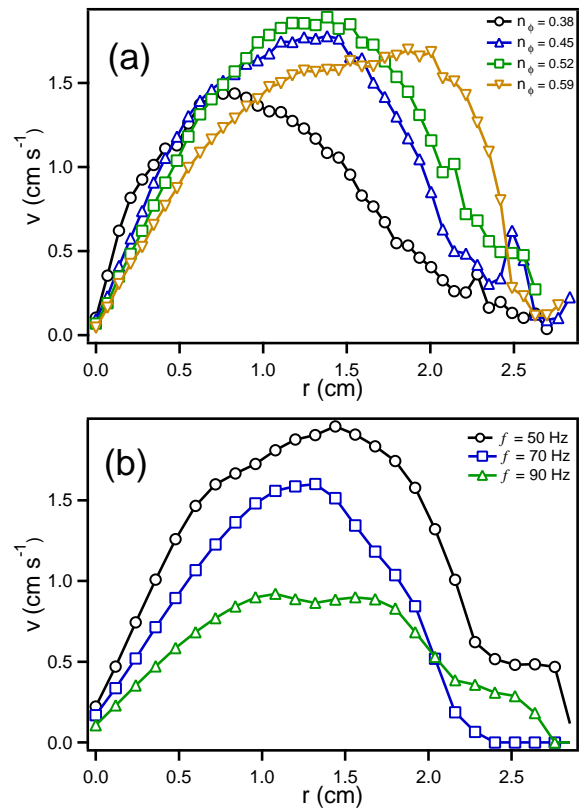}
\caption{\label{Fig4} (a) The azimuthal averaged velocity $v(r)$ as a
function of  the distance $r$ from the center of the vortex. The
velocity is observed to increase linearly from the center of the
vortex and then reach a maxima before decreasing to the edge of the
vortex. The rate of increase of velocity at the center is observed to
systematically decrease with $n_\phi$ and hence vortex size. (b)
$v(r)$ as function of frequency $f$ for a fixed $\Gamma$ at
$n_{\phi}=0.53$.  The velocity of rotation is observed to depend on
the velocity of the vibration. $\Gamma = 3.00$ for all data shown. }
\end{figure}

To describe the nucleation and growth of the vertical domains at high
$n_{\phi}$ shown in Fig.~\ref{vort_t} and \ref{coarsen}  we present a
simple analysis.  The rods are assumed to be   either vertical or
horizontal.  We then assume that the growth rate is initially linear
and asymptotically must decrease to zero as the number of horizontal
rods is diminished and a steady state is reached.   Then the evolution
of $n_v$ is described by,

\begin{equation}
\frac{\partial n_v}{\partial t} = \alpha n_v( \beta - n_v).
\label{model1}
\end{equation}
where, $\alpha$ and $\beta$ are constants that depend on $n_\phi$.
Eq.~\ref{model1} can be integrated and is fit to the measured $n_v(t)$
in Fig.~\ref{coarsen}.  This simplified interpretation seems to
capture the nucleation, growth, and saturation of the near-vertical
domains in our experiments.

In previous work, Villarruel {\em et al.}~\cite{Villarruel}, the
experiments were carried out in a system where the diameter of the
container and the length of the rods were comparable. The container
was tapped and thus the rods experienced considerable shearing with
the side boundary.  They observed that vertical domains nucleated at
the boundaries and subsequently propagate to the center of the cell.
These observations are consistent with our findings.  We have shown
that vertical domains can nucleate away from the sidewalls and form
independent of the direction of the driving.

Next, we elucidate the physical mechanism that is responsible for the
vortex motion.  We performed additional experiments with a row of
cylindrical rods with $l =$ 5.1~cm and $d =$ 0.6~cm in a 1-D annulus
with a mean radius of 5.5~cm [see Fig.~\ref{annu}].  When subjected to
vertical motion, the rods were always observed to move in the
direction of the inclination. No translation motion is observed when
the rods were vertical. A movie showing this property can be found at
Ref.~\cite{movies}. We also used ball point pens and pencils to check
that the detailed shape of the rod and the tip is not important to
this translation mechanism. We find that the translation speed depends
on the driving frequency and the inclination, qualitatively similar to
that observed for the vortices.

The physical mechanism for the motion of the inclined rods, based on
our observation, appears to be as follows.  When the inclined rods are
vibrated vertically, they hit the bottom plate at a point away from
their center of mass.  Because the rotation of the rod about its
center of mass is constrained due to the neighboring rods, it gets
launched in the direction of the inclination. Thus, the greater the
inclination of the rods, the further they get launched and land ({\em
viz.}~projectile motion), giving rise to translational motion.

From these direct observations we conclude that the inclined rods form
the engine that drives the vortex.  The vertical rods in the center of
the vortex are simply pulled around by shear induced by the inclined
rods. As the number of rods is increased and the size of the vortex
increases, the inclination of the rods also decreases because of
greater packing. Therefore, the speed of the very large vortices is
expected to decrease with the size of the vortex, consistent with our
observation [see Fig.~\ref{Fig4}(a)].  We also note that the frequency
dependence of $v(r)$  is compatible with the mechanism described
above.  By decreasing the frequency of the driving signal at fixed
$\Gamma$ the velocity of the driving plate is increased.  Therefore,
one expects $v(r)$ to increase with decrease  in frequency, consistent
with our observations [see Fig.~\ref{Fig4}(b)].
\begin{figure}[t] 
\includegraphics[width=0.4\textwidth]{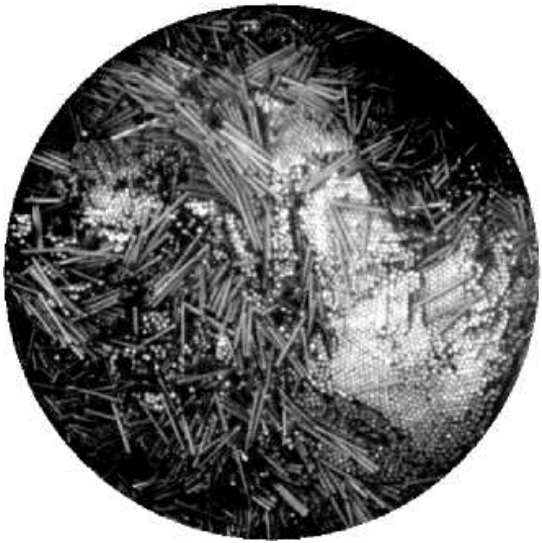}
\caption{\label{horizontal}Image of domains of vertical rods formed
when the container is vibrated horizontally.  We note that vortices
are not observed.  The direction of vibration is from left to
right. $\Gamma = 1.04$, $f = 30$ Hz, and $n_{\phi} = $ 0.612. }
\end{figure}
\begin{figure}[t] 
\includegraphics[width=0.4\textwidth]{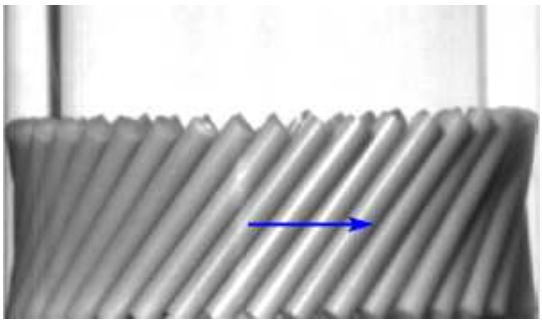}
\caption{\label{annu} Image of rods contained in a vertically vibrated 1-D
annulus. The rods are always observed to move in the direction of
inclination. See text and Ref.~\cite{movies} for more details.}  
\end{figure}

\section{Conclusion}
In conclusion, granular rods are observed to form vertically aligned
domains at high packing fractions, independent of the direction of
vibration. Novel vortex patterns are observed when the rods are
vibrated vertically.  We have measured the growth of near-vertical
domains and have developed a simple {\em void filling} model that well
describes our results.  We have also shown a new translation mechanism
which occurs due to anisotropy.  Based on these observations, Aranson and
Tsimring~\cite{aranson} have developed a phenomenological model that
describes the formation and coarsening of the vortices.

\begin{acknowledgments}
We acknowledge stimulating discussions with Lev Tsimring and Igor
Aranson, and Eric Frederick for assistance with the experiments. 
We thank Seth Fraden for inspiring the experiments, for
sharing unpublished data, and for providing us with some of the rods
which were purchased under his NSF grant, DMR-0088008. This work was
supported by the NSF under Grant No. DMR-9983659, and by the donors of
the Petroleum Research Fund.
\end{acknowledgments}


\begin{thebibliography}{}

\bibitem{Melo94} F. Melo, P. B. Umbanhowar and H. L. Swinney,
Phys. Rev. Lett. {\bf 72}, 172 (1994); {\it ibid} {\bf 75}, 3838
(1995).

\bibitem{knight93} J. B. Knight, H. M. Jaeger and S. R. Nagel,
Phys. Rev. Lett. {\bf 70}, 3728 (1993).

\bibitem{tennakoon98} S. G. K. Tennakoon and R P Behringer,
Phys. Rev. Lett. {\bf 81}, 794 (1998).

\bibitem{makse97} H. A. Makse, S. Havlin, P. R. King and
H. E. Stanley, Nature (London) {\bf 386}, 379 (1997).

\bibitem{Mounfield} C. C. Mounfield and S. F. Edwards, Physica A {\bf
210}, 279 (1994).

\bibitem{Villarruel} F. X. Villarruel, B. E. Lauderdale, D. M. Mueth
and H. M. Jaeger, Phys. Rev. E {\bf 61}, 6914 (2000).

\bibitem{Onsager} L. Onsager, Ann. N.Y. Acad. Sci. {\bf 51}, 627
(1949).

\bibitem{Khokhlov} V. A. Baulin and A. R. Khokhlov, Phys. Rev. E {\bf
60}, 2973 (1999); B. J. Buchalter and R. M. Bradley,
Europhys. Lett. {\bf 26}, 159 (1994).

\bibitem{Dogic} Z. Dogic and S. Fraden, Phys. Rev. Lett. {\bf 78},
2417 (1997).

\bibitem{Kooij}  F. M. van der Kooij, K. Kassapidou and
H. N. W. Lekkerkerker, Nature {\bf 406}, 868 (2000).

\bibitem{deGennes} P. G. deGennes and J. Prost, {\it The Physics of
Liquid Crystals}, (Oxford Science, Second Edition, 1993).

\bibitem{heap-ref} P. Evesque and J. Rajchenbach,
Phys. Rev. Lett. {\bf 62}, 44 (1989).

\bibitem{fraden} S. Fraden, private communication.

\bibitem{movies} The movies showing the vorticity of the patterns can
be viewed at: \url{<http://physics.clarku.edu/vortex>}.

\bibitem{IDL} J. C. Crocker and D. G. Grier, J. Colloid Interface
Sci. {\bf 179}, 298 (1996).

\bibitem{aranson} I. S. Aranson and L. S. Tsimring, \eprint{cond-mat/0203237}.

\end{thebibliography}
\end{document}